\documentclass[aps,pre,onecolumn,showpacs,a4paper]{revtex4}
\usepackage{graphicx}
\usepackage{amsmath}
\usepackage{amssymb}
\usepackage{amscd}
\usepackage[dvips]{epsfig}
\newcommand{\clA}{{\cal A}}

\newcommand{\clW}{{\cal W}}

\newcommand{\prt}{\partial}

\newcommand{\rgl}{\rangle}
\newcommand{\lgl}{\langle}

\newcommand{\factor}{\mbox{!}}

\begin{document}
\title{Asymptotic localization of stationary states in the nonlinear
Schr\"odinger equation}

\author{Shmuel Fishman${}^a$, Alexander Iomin${}^a$, and Kirone
Mallick${}^b$}

\affiliation{${}^a$ Department of Physics, Technion, Haifa, 32000,
Israel \\
${}^b$ Service de Physique Th\'eorique, Centre d'\'Etudes de Saclay, 91191
Gif-sur-Yvette Cedex, France}

\date{\today}
\begin{abstract}
The mapping of the Nonlinear Schr\"odinger Equation with a random 
potential on the Fokker-Planck equation is used to calculate the 
localization length of its stationary states. The asymptotic growth rates
of the moments of the wave function and its derivative for the linear 
Schr\"odinger Equation in a random potential are computed analytically and
resummation is used to obtain the corresponding growth rate for the 
nonlinear Schr\"odinger equation and the localization length of the 
stationary states.

\end{abstract}


\maketitle

\hfill\break
\section{Introduction  }

In this work we consider the problem of the one-dimensional
Anderson localization \cite{Anderson,Lee} for the nonlinear
Schr\"odinger equation (NLSE) \cite{bwFSW,bwDS,kivshar}. In spite
of the extensive research, many fundamental problems are still
open, and, in particular, it is not clear whether in one dimension
(1D) Anderson localization can survive the effects of
nonlinearities. This problem is relevant for experiments in
nonlinear optics, for example disordered photonic lattices
\cite{f1,lahini}, where Anderson localization was found in presence of
nonlinear effects as well as experiments on Bose-Einstein
Condensates (BEC) in disordered optical lattices
\cite{BECE1,BECE3,akkermans,SanchAspect,BShapiro}. The interplay
between disorder and nonlinear effects leads to new interesting
physics \cite{BECE3,akkermans,bishop1,aubry,mackay,aub1}. In
particular, the problem of spreading of wave packets and
transmission are not simply related \cite{doucot, pavloff}, in
contrast with the linear case.

We consider one-dimensional  localization of stationary solutions
of the nonlinear Schr\"odinger equation (NLSE) in a random
$\delta$-correlated potential $V(x)$ with a Gaussian distribution (white 
noise), of zero mean and variance  $2D$, namely, 
$\lgl V(x)V(x')\rgl=2D\delta(x-x')$.
The linear version of this model
was studied extensively in the past \cite{LGP}. Following our
previous analysis \cite{if2007} we study Anderson localization of
stationary solutions with energies $\omega$ in the framework of
the stationary NLSE
\begin{equation}\label{nlse_A1}
\omega\phi(x)=-\prt_{x}^2\phi(x)+\beta\phi^3(x)+V(x)\phi(x)\, ,
\end{equation}
where $\phi(x)$ is real, the variables are chosen in
dimensionless units and the Planck constant is $\hbar=1$. For the
lattice version of the model, it was established rigorously
\cite{bwFSW,AF1,AF2} that the stationary solutions of this
equation are exponentially localized for a wide range of
conditions. It  was  also  argued that the rate of growth of moments for
the stationary NLSE (\ref{nlse_A1}) coincides exactly with the
linear case \cite{if2007} and determines the localization length.

We will specifically calculate $\lgl\phi^2(x)\rgl$ of solutions of
Eq. (\ref{nlse_A1}) that are found for a certain $\omega$, with
given boundary conditions at some point, for example $\phi(x=0)$
and $\phi^{\prime}(x=0)$, where prime means the derivative with
respect to $x$. This will be done with the help of the analogy with the 
Langevin equation \cite{LGP,if2007,halperin}. In particular, we will 
calculate the growth rate of the second moment,
\begin{equation}\label{nlse_A2}
2\gamma=\lim_{x\to\infty}\frac{ \ln\lgl\phi^2(x)\rgl}{x}>0\, ,
~~~\xi=\frac{1}{\gamma}\, ,
\end{equation}
that will turn out to be independent of $\beta$, where $\xi$ is the 
localization
length. Note, it is different from the usually studied (in the
linear case) self averaging quantity
$\gamma_s=\frac{1}{2}\frac{d}{dx}\lgl\ln\phi^2(x)\rgl=
\frac{1}{2}\lim_{x\to\infty}\frac{\lgl\ln\phi^2(x)\rgl}{x}$, and $\gamma$
is a smooth function of energy. Since the distribution of the random
potentials is translationally invariant, it is independent of the
choice of the initial point as $x=0$. Like in the linear case,
starting from a specific initial condition, $\phi(x)$ will
typically grow. For specific values of $\omega$ at some point this
function will start to decay, so that a normalized eigenfunction
is found. This is the approach  {\it \`a la} Borland 
\cite{borland,mott} that was made rigorous for the linear case in 
\cite{delyonA,delyonB}. 
Here, following \cite{if2007} we extend this approach, in a heuristic form, 
to the nonlinear case. The envelope of the
wave function will grow exponentially if we start either from the
right or from the left. The value of $\omega$ results from the
matching condition, so that an eigenfunction has some maximum and
decays in both directions as required by the normalization
condition. The exponential decay is an asymptotic property, while
the matching is determined by the potential in the vicinity of the
maximum. This observation is crucial for
the validity of this approach and enables us to determine the
exponential decay rate of states from the solution of the initial
value problem (\ref{nlse_A1}). 
In \cite{if2007} a linear equation for the moments of $\phi(x)$ of Eq. 
(\ref{nlse_A1}) and its Langevin analog was derived and it was shown 
that the exponents that control the growth of the moments are identical to 
the ones of the linear system $(\beta=0)$. 
In the present paper it will be shown that this is also correct 
for the asymptotic behavior of the moments.

The outline of the paper is as follows. The analogy with the Langevin 
equation and dynamics of the moments is outlined in Section II. The 
generalized Lyapunov exponents, that are the eigenvalues which 
determine the growth of the various moments, are presented in Section II
and their asymptotic behavior is derived analytically for the first time 
in Appendix A (they were 
found numerically in \cite{pikovsky}). The resulting asymptotic expansion 
for the growth of the moments is presented in Section III. The results are 
summarized in Section IV. 

\section{The Fokker-Planck equation and Lyapunov exponents}

Following Ref. \cite{if2007}, we perform the calculation of
$\lgl\phi^2(x)\rgl$ by using the analogy with the classical
Langevin equation \cite{LGP,halperin}. Therefore,  considering the
$x$-coordinate as the formal time variable on the half axis
$x\equiv\tau\in[0,\infty)$,  Eq. (\ref{nlse_A1}) reduces to the
Langevin equation
\begin{equation}\label{nlse4}
\ddot{\phi}+\omega\phi-\beta\phi^3-V(\tau)\phi=0
\end{equation}
with the $\delta$ correlated Gaussian noise $V(\tau)$. Now we
introduce new variables $u=\phi$ and $v=\dot{\phi}\equiv
\frac{d\phi}{d\tau}$ (that play the role of position and velocity in the 
Langevin equation) and a distribution function 
of these new
variables is $P=P(u,v,\tau)$. The dynamical process in the
presence of the Gaussian $\delta$-correlated noise is described by
the distribution function that satisfies the Fokker-Planck
equation: (FPE) \cite{if2007,kmpmarcq}
\begin{equation}\label{nlse5}
\prt_{\tau}P-[\omega u-\beta u^3]\prt_vP+v\prt_uP-2D u^2
\prt_{v}^2P=0\, ,
\end{equation}
where $2Du^2$ is the only nonzero component of the diffusion
tensor.

\subsection{Equation for moments}

We are interested in the average quantum probability density
$\lgl\phi^2(x)\rgl\equiv \lgl u^2(\tau)\rgl$, where
\[\lgl u^2(\tau)\rgl=\int u^2P(u,v,\tau)dudv\, .\]
From the FPE we obtain a system of equations for the moments
\begin{equation}\label{nlse6}
M_{k,l}=\lgl u^kv^l\rgl\, ,
\end{equation}
where $k,l=0,1,2,\dots$. Substituting $u^kv^l$ in the FPE and
integrating over $u$ and $v$, one obtains the following relation
for $M_{k,l}$
\begin{equation}\label{nlse7}
\dot{M}_{k,l}=-l\omega M_{k+1,l-1}+kM_{k-1,l+1} +l(l-1)2D
M_{k+2,l-2} +\beta l M_{k+3,l-1}\, ,
\end{equation}
where $ M_{k,l}$ with negative indexes are assumed to vanish. We
note that only terms with the same parity of $k+l$ are coupled.
Since we are interested in $M_{2,0}=\lgl u^2\rgl$, we study only
the case when this parity is even, namely $k+l=2n$ with
$n=1,2,\dots$.
  The sum of the indexes of the moments
is $2n$, except the last term $\beta l M_{k+3,l-1}$, where the sum
is $2(n+1)$. This leads to the infinite system of linear equations
that can be written in the form
\begin{equation}\label{nlse8}
\dot{\bf M}=\clW{\bf M}\, ,
\end{equation}
where the column vector is ${\bf
M}=\left(M_{2,0},M_{1,1},M_{0,2},M_{4,0},M_{3,1},\dots\right)$ and
$\clW$ is the corresponding matrix. The matrix elements
$\clW_{k,l}$ are determined by Eq. (\ref{nlse7}). The solutions of
the system of linear equations (\ref{nlse8}) are linear
combinations of the eigenfunctions at time $\tau$
\begin{equation}\label{nlse9}
{\bf M}_{\lambda}(\tau)=\exp(\lambda\tau){\bf U}_{\lambda}\, ,
\end{equation}
where ${\bf U}_{\lambda}={\bf M}_{\lambda}(t=0)$ is the
eigenvector of $\clW$ corresponding to the eigenvalue $\lambda$
found from the equation
\begin{equation}\label{nlse13}
\clW {\bf U}_m=\lambda_m {\bf U}_m\, , ~~ ~{\bf U}_m\equiv {\bf
U}_{\lambda_m}\, .
\end{equation}
The infinite matrix $\clW$ consists of two parts. The first one is
independent of $\beta$ and consists of independent diagonal blocks
$A_n$  of size $(2n+1)\times (2n+1)$. The second one consists of
the $\beta$ dependent terms which couple the $n$-th  and  the
$(n+1)$-th blocks and are located above the $(n+1)$ block and to
the right of the $n$-th block. Consequently, the $\beta$ dependent
terms do not affect the characteristic polynomial, as can be shown
by elementary operations on determinants. Therefore, the
characteristic polynomial of $\clW$ reduces to a product of the
block determinants \cite{if2007}
\begin{equation}\label{nlse10}
\prod_{n=1}^{\infty}{\rm det}\left(A_n-\lambda I_n\right)=0\, ,
\end{equation}
where $I_n$ is an $(2n+1)\times(2n+1)$ unit matrix. The diagonal block 
$A_n$ of  the infinite matrix $\clW$ defined in equation~(\ref{nlse8}),
that couples the moments of order $2n$ with one another, is a 
band-diagonal square-matrix of size $(2n+1)$.
The explicit form of this matrix is given by:
\begin{equation}
A_n  =  \left( \begin{array}{ccccccccc} \quad \quad  0 \quad
\quad   &\quad \quad 2n  \quad  \quad   &
 \quad  \quad   0  \quad  \quad  &  \quad  \quad & \quad  \quad
  & \quad  \quad & \quad  \quad  & \quad  \quad & \\
 -\omega & 0  & 2n-1 &  0 &  & &  \\
 2  D&  -2\omega & 0  & 2n-2 & 0 & &   \\
   0  &  6 D&  -3\omega & 0  & 2n-3& 0&   \\
       &   &  . & . & . &  .  & \\
  &   &   &.  & . &  . &  .  &   \\
  &   &   &  & . &  . &  .  &  . \\
         &  &    &  &  0 &  (2n-2)(2n-1) D \quad  \quad&
    -(2n-1)\omega & 0  & 1 \\
         &  &     &   &  & 0 &   (2n-1)2n D& -2n \omega & 0
\end{array}  \right)
\label{formulaAn}
\end{equation}
Let us denote by $\lambda_{{\rm max}}(n)$ the maximal eigenvalue of this 
matrix. The vector of the moments in this block is ${\bf M}_n=
(M_{2n,0},M_{2n-1,1}\dots, M_{2n-l,l},\dots,M_{0,2n})$.
In Appendix A it is proven that for $\omega=0$ the maximal Lyapunov 
exponent $\lambda_{\rm max}(n)$ behaves for large $n$ as 
 \begin{equation}
  \lambda_{{\rm max}}(n)  \simeq \frac{3}{4} D^{1/3}  (2n)^{4/3}  \, .
\label{eq:scalingIIB}
 \end{equation}
Then it is argued and verified numerically that also for other values 
$\omega\neq 0$ it behaves in this way.

\section{Asymptotic growth of the moments}

\subsection{Eigenvalue problem for the moments}

Taking into account Eqs. (\ref{nlse9}) and (\ref{nlse13}), we present
the solution of Eq. (\ref{nlse8}) as an expansion
\begin{equation}\label{asy1}
{\bf M}(\tau)=\sum_mC_m(\tau){\bf
U}_m=\sum_me^{\lambda_m\tau}c_m{\bf U}_m\, ,
\end{equation}
where $c_m\equiv C_m(\tau=0)$. Due to the block structure the
eigenvectors are characterized by two indexes $m=(n,k)$, where $n$
indicates number of block, while  $k=1,2,\dots\, , 2n+1$  counts
elements
inside each block. Therefore, the eigenstates  
${\bf U}_m\equiv{\bf U}_{n,k} $ 
are found from the following algorithm. For the block $n=1$ there are
3 eigenvalues $\lambda_{1,k}$ with corresponding eigenvectors
$\bar{\bf U}_{1,k}$ determined by the first block $A_1$.
Therefore, ${\bf U}_{1,k}=(\bar{\bf U}_{1,k}, {\bf O})$, where
${\bf O}$ is an infinite zero vector. For the second block $n=2$ there are
5 eigenvalues $\lambda_{2,k}$ with corresponding eigenvectors ${\bf
U}_{2,k}=({\bf R}_{1,k},\bar{\bf U}_{2,k},{\bf O})$, where ${\bf
R}_{1,k}$ is a 3 dimensional vector, while $\bar{\bf U}_{2,k}$ is
a 5 dimensional vector, and $k=1,2,\dots,5$. Here $\lambda_{2,k}$
and $\bar{\bf U}_{2,k}$ are determined from the second diagonal
block matrix $A_2$, while ${\bf R}_{1,k}$ is determined by $A_1$
and by the  corresponding $\beta$-dependent  off diagonal block.
Continuing this procedure, we obtain $2n+1$ eigenvectors for
$\lambda_{n,k}$ in the form
\begin{equation}\label{asy2}
{\bf U}_{n,k}=({\bf R}_{n-1,k},\bar{\bf U}_{n,k},{\bf O})\, ,
\end{equation}
where ${\bf R}_{n-1,k}$ is a $(n^2-1)$ dimensional vector
determined by $n-1$ diagonal and off diagonal blocks of the
truncated matrix $\clW$.

Summation over $m$ in Eq.
(\ref{asy1}) is broken into the sum over
the block numbers $n\in[1,\infty)$ and the sum over indexing
inside each block $l\in[0,2n]$. Thus, Eq. (\ref{asy1}) reads
\begin{equation}\label{asy3}
{\bf M}(\tau)=\sum_{n=1}^{\infty}\sum_{l=0}^{2n}
c_{n,l}e^{\lambda_{n,l}\tau}{\bf U}_{n,l}\, .
\end{equation}
The vector ${\bf M}$ consists of the block vectors ${\bf M}_n$:
${\bf M}=({\bf M}_1,{\bf M}_2,\dots\, , {\bf M}_n,\dots)$, where
  ${\bf M}_n$ is a vector of $2n+1$ elements defined in Eq.
(\ref{nlse6}) and corresponds to the moments of the order of $2n$.
Therefore, the initial vector at $\tau=0$ is
\begin{equation}\label{asy4}
({\bf M}_1,{\bf M}_2,\dots\, , {\bf M}_n,\dots)=
\sum_{n'}\sum_{l'}c_{n',l'}{\bf U}_{n',l'}\, .
\end{equation}
Assume that at the initial point $\tau=0$, the wave function and
its derivative are small, of the order of $\epsilon$
 in units  of $\sqrt{|\omega/\beta|}$. Then, at
that point the moments scale as ${\bf M}_n\sim\epsilon^{2n}$ with 
$\epsilon$ arbitrary small. 
As follows from Eq. (\ref{asy4}), one finds $c_{n,l}U_{n,l}\sim
\bar{c}_n\epsilon^{2n}+o(\epsilon^{2n})$ with bounded $\bar{c}_n$
for any $n$, as demonstrated in Appendix B.
In the linear case $(\beta=0)$ the growth rate of each
moment of the order of $2n$ corresponding to the $n$-th block is
determined by the eigenvalue with the largest real part, and we
denote it by $\lambda_{\rm max}(n)\equiv\max({\rm Re\,}\lambda_{n,l}) $, 
where the maximum is over the $2n+1$ eigenvalues corresponding to the 
$n$th block, indicated by $l$. As shown in 
Appendix \ref{sec:Appendix1}, the asymptotic behavior of the
generalized Lyapunov exponent  $\lambda_{\rm max}(n)$ when  $n \to \infty$
is given by  $\lambda_{\rm max}(n)\sim\clA n^{4/3}$, where 
$\clA=\frac{3}{4}2^{4/3}D^{1/3}$ is a
constant (see Eq. (\ref{eq:scalingIIB})). The leading contribution to the 
growth of ${\bf M}_n(\tau)$ in 
the nonlinear case $(\beta\neq 0)$ is determined by the following sum
\begin{equation}\label{asy5}
\tilde{\bf M}_n(\tau)= \sum_{m\ge n}\bar{c}_m\epsilon^{2m}e^{\clA
m^{4/3}\tau}\, ,
\end{equation}
as is clear from Eqs. (\ref{asy1})--(\ref{asy4}).

\subsection{Resummation}

This series in Eq. (\ref{asy4}) has a vanishing  radius of
convergence and probably has to be interpreted as an asymptotic series. 
It can be used to study  the behavior of $\tilde{\bf M}_n(\tau)$ after 
being  
resummed. Such a resummation is done with the help of the identity
\begin{equation}
\exp(K^2) = \frac{1}{\sqrt \pi}
\int_{-\infty}^{+\infty}  du  \exp(-u^2 + 2Ku)\, ,
\label{HubbStrato}
\end{equation}
known as the Hubbard-Stratonovich transformation. We can  rewrite the 
above  series as follows
\begin{equation}\label{HST}
\tilde{\bf M}_n(\tau)=\sum_{m\ge n} \bar{c}_m \epsilon^{2m} \exp( \clA 
m^{4/3}\tau) = \frac{1}{\sqrt \pi}
 \int_{-\infty}^{+\infty}  du  \exp(-u^2) \Phi(2u\sqrt{ \clA\tau})\, ,
\end{equation}
 where the function  $\Phi$ is given by
\begin{equation}
 \Phi(y) = \sum_{m\ge n}  \bar{c}_m \epsilon^{2m} \exp(m^{2/3}y) \, .
\end{equation}
 If the coefficients  $\bar{c}_m$ do not grow too fast (which we shall assume
 hereafter)  the function $  \Phi(y)$ is well defined at least for small 
values of $y$. Note that for the resummation of Eq. (\ref{HST}) it was not 
crucial that the power of $m$ is $4/3$. Such a resummation can be 
performed for any power $\bar{\alpha}<2$ replacing $4/3$.

A more explicit resummation procedure for the r.h.s. of 
equation~(\ref{asy5})
 can  be developed with the help of fractional derivatives.
First, let us expand the exponential function
\begin{equation}\label{asy5b}
\exp(\clA\tau m^{4/3})=\sum_{k=0}^{\infty}\frac{\left(\clA\tau
m^{4/3}\right)^k}{k\factor}\, .
\end{equation}
Then, writing $\epsilon^{2m}=e^{m\ln\epsilon^2}\equiv e^{\zeta m}$, we 
obtain
\begin{equation}\label{asy5c}
\tilde{\bf M}_n(\tau)=
\sum_{k=0}^{\infty}\frac{\left(\clA\tau\right)^k}{k\factor}
\sum_{m\ge n}\bar{c}_m m^{4k/3}e^{\zeta m}\, .
\end{equation}
We now  introduce the Weyl fractional derivative of order $q$ of a function
 $f(z)$ by the Weyl integral, see e.g.  \cite{SokKlafBlum},
\begin{equation}\label{asy5e}
\frac{d^q f(z)}{dz^q}\equiv
\frac{1}{\Gamma(-q)}\int_{-\infty}^z\frac{f(y)dy}{(z-y)^{1+q}}
\end{equation}
 where  for $q>0$ the integral should be properly regularized 
\cite{SokKlafBlum,gelfand}, and $\Gamma(-q)$ is the gamma function.
For $f(y)=e^{\nu\zeta}$ it takes the form 
\begin{equation}\label{asy5d}
\frac{d^qe^{\nu\zeta}}{d\zeta^q}=\nu^qe^{\nu\zeta}\, .
\end{equation}
Substitution of Eq. (\ref{asy5d}) in Eq. (\ref{asy5c}) with $\nu=m$
yields
\begin{equation}\label{asy5f}
\tilde{\bf M}_n(\tau)=
\sum_{k=0}^{\infty}\frac{\left(\clA\tau\right)^k}{k\factor}
\frac{d^{4k/3}}{d\zeta^{4k/3}}\sum_{m\ge n}\bar{c}_m\epsilon^{2m}\, .
\end{equation}
If $\bar{c}_m$ are bounded, as shown in Appendix B, by some $\bar{C}_n$, 
the sum in Eq. 
(\ref{asy5f}) is bounded by $\frac{\bar{C}_ne^{2n}}{1-\epsilon^2}$, 
hence $\tilde{\bf M}_n$ is the fractional derivative of some well defined 
function presented in Eq. (\ref{asy5e}) with its regularization. 

Therefore, Eqs. (\ref{asy5f}) and (\ref{HST}) describe the long time 
asymptotics of the moments. Therefore, Eq. (\ref{asy5}) 
is an asymptotic expansion and is a good approximation as long as we 
sum decreasing terms. The condition is (for bounded $\bar{c}_m$)
\begin{equation}
\epsilon^{2m}\exp\left(\clA m^{4/3}\tau\right) >
\epsilon^{2(m+1)}\exp\left(\clA (m+1)^{4/3}\tau\right)\, .
\end{equation}
For large $m$ this inequality is 
$\ln\frac{1}{\epsilon}>\frac{2}{3}\clA\tau 
m^{1/3}$ with $m\geq n$.    
Consequently, for the time of the order
\begin{equation}\label{asy6}
\tau< \tau_0^{(n)}\equiv\frac{3}{2}\frac{\ln(1/\epsilon)}{\clA n^{1/3}}
\end{equation}
the $n$-th moment will be dominated by the leading terms and will grow as 
in the linear case $(\beta=0)$.

\subsection{The growth of the second moments}

The second moments are of particular interest for the present work. Their 
growth for a time that is shorter than $\tau_0^{(1)}$ of Eq. (\ref{asy6}) 
is dominated by the leading term, namely $\tilde{\bf M}_2=
\bar{c}_2\epsilon^2e^{\lambda_{\rm max}(1)\tau}$. Consequently, for 
$\tau<\tau_0^{(1)}$, 
 \begin{equation}
 \lgl u^2\rgl=M_{2,0}=\bar{c}_2\epsilon^2e^{\lambda_{\rm{max}}(1)\tau}\,
.\end{equation}
This result was verified numerically. Using the analogy between the 
stationary Schr\"odinger equation (\ref{nlse_A1}) and the Fokker-Planck 
equation (\ref{nlse5}) we identify the generalized Lyapunov exponent
$\lambda_{\rm max}(1)$ with the growth rate (\ref{nlse_A2}) as 
\begin{equation}\label{nlse_A2b}
2\gamma=\lambda_{\rm max}(1)  =\lim_{x\to\infty} \lim_{\epsilon\to 0}
\frac{ \ln\lgl\phi^2(x)/\epsilon^2 \rgl}{x}>0\, .
\end{equation}
 For $\tau>\tau_0^{(1)}$,  non-linear saturation  effects become relevant
 and the full nonlinear theory should be used.

\section{Summary}

In this work the asymptotic behavior of the generalized Lyapunov exponents 
of the linear Fokker-Planck equation (\ref{nlse4}) with $\beta=0$, was 
found analytically and is given by Eq. (\ref{eq:scalingIIB}). The 
resulting 
expression for the moments (\ref{asy5}) is divergent, but it can be 
resummed in the form Eqs. (\ref{HST}) or (\ref{asy5f}). Therefore, for 
short time $\tau<\tau_0^{(n)}$, the first term in Eq. (\ref{asy5}) 
provides 
a good approximation of the moments. In particular, for the second moment 
this result enabled to identify the generalized Lyapunov exponent of the 
linear system with the asymptotic growth rate for the nonlinear one, 
namely 
\[2\gamma=\lambda_{\rm max}(1)\, .\]
According to the implementation of the \textit{\`a la} Borland method, as 
outlined in \cite{if2007}, this is the decay rate of the stationary states 
of the random nonlinear equation (\ref{nlse_A1}), 
showing that it is independent of the nonlinearity $\beta$.

\subsection*{Acknowledgments}
We thank Arkady Pikovsky for instructive comments and criticism of
\cite{if2007}. This work was supported in part by the Israel Science
Foundation (ISF), by the US-Israel Binational Science Foundation (BSF),
by the Minerva Center for Nonlinear Physics of Complex Systems, the Fund 
for encouragement of research at the Technion and Shlomo Kaplansky academic 
chair.

\appendix

\section{Asymptotic behavior of the generalized Lyapunov exponents}
\label{sec:Appendix1}

In this appendix, $\lambda_{{\rm max}}(n)$, the maximal eigenvalue
 of the matrix (\ref{formulaAn}) will be evaluated. Following  
\cite{pikovsky} we call this  quantity a  {\it generalized Lyapunov 
exponent}.
 By elementary dimensional analysis \cite{pikovsky}, one finds that the 
following scaling is suggestive,
\begin{equation}
 \lambda_{{\rm max}}(n) = D^{1/3} L_n\left(  \frac{\omega}{D^{2/3}} 
\right)
  \, .
 \end{equation}
  In the long  time limit
 the moments of order $2n$  grow  as $e^{ \lambda_{{\rm max}}(n) t }$.
 In  \cite{pikovsky},  Zillmer and Pikovsky studied
 $\lambda_{{\rm max}}(n)$ for different values  of $n$:  
  they give  exact expressions for $n=2$,  for
 $n \to 0$ (which corresponds to the usual Lyapunov exponent)   
 and consider  limiting  cases for  large values of the dimensionless
   parameter
 ${\omega}/{D^{2/3}}$.  They also study  numerically the behavior
 of  $\lambda_{{\rm max}}(n)$ as $ n \to \infty$, keeping
 the dimensionless parameter fixed.   They found numerically the
 scaling law:
 \begin{equation}
 \lambda_{{\rm max}}(n) \propto  n^{\alpha}
 \,\,\,\, \hbox{ with } \,\,\, \alpha \simeq 1.4 \, .
\label{largenscaling}
\end{equation}
 The fact that the scaling  exponent $\alpha$  is different
 from 2, implies deviations from  Gaussian behavior
 and breakdown of monoscaling; the consequences of this  breakdown of 
single
 parameter scaling   for   conductance distribution
  were  studied by Schomerus and Titov \cite{shomerus}.
 We also remark that  related studies  were  also carried out
 in the  context of the
  harmonic oscillator with random frequency   \cite{lindenberg,kmpmarcq}.

In this appendix, we study  analytically the  behavior
of the generalized Lyapunov exponents in the limit $n \to \infty$.
 For the special case of  $\omega=0$, we prove   the
 following asymptotic formula:
 \begin{equation}
  \lambda_{{\rm max}}(n)  \simeq \frac{3}{4} D^{1/3}  (2n)^{4/3}  \, .
\label{eq:scaling}
 \end{equation}
 We shall then  argue that this behavior remains valid for any finite 
value
 of $\omega$.

 We now outline the proof of the scaling
 equation~(\ref{eq:scaling}) for   $\omega=0$  by   
 studying the large $n$ behavior of the
    coefficients of the   characteristic polynomial $P(X)$
 of the matrix $A_{n}$.  Recalling   that if $\lambda_{n,1}, \ldots
  \lambda_{n,2n+1}$  are the  eigenvalues of $A_{n}$  we have
 \begin{equation}
    P(X) = \prod_{i=1}^{2n+1} (X -\lambda_{n,i})  = X^{2n+1} -
  (\sum_{i=1}^{2n+1} \lambda_{n,i}) X^{2n} +
  (\sum_{i\neq j} \lambda_{n,i} \lambda_{n,j}) X^{2n-1} + \ldots +
  \prod_{i=1}^{2n+1} \lambda_{n,i}  \, .
\end{equation}
We also have $\lambda(n) =\max_{l}({\rm Re\,}\lambda_{n,l})$.
 The   coefficients of the   characteristic polynomial $P(X)$
 are symmetric functions of the eigenvalues $\lambda_{n,i}$.
 Thanks to the  Newton formulae,  all
 these   coefficients   can be written
 as linear combinations of traces  of powers of   $A_{n}$. Hence, we have
 \begin{eqnarray}
    P(X)
   = X^{2n+1} - {\rm Tr}(A_{n})  X^{2n} + \frac{1}{2}
 \Big( [{\rm Tr} (A_{n})]^2 -{\rm Tr}(A_{n} ^2) \Big)
  X^{2n-1} + \ldots
\end{eqnarray}
In principle, we can obtain the eigenvalues from the traces of the $2n+1$ 
powers of $A_n$. In practice, since we are interested only in the 
asymptotic behavior of $\lambda_{\rm max}(n)$, it will be computed from 
the traces of high powers.
   From equation~(\ref{formulaAn})
 we observe that   ${\rm Tr}(A_{n}) = 0$. In the case
  $\omega=0$,  we also have  ${\rm Tr}(A_{n}^2) = 0$.
  More generally,  for  $\omega=0$,
we can show that ${\rm Tr}(  A_{n}^k) \neq  0$  only when
  $k$ is a multiple of 3.
 Indeed, writing
 \begin{equation}
     A_{n}= d + g  
  \label{eq:AnasSum}
\end{equation}
 with
 \begin{equation}
  d = \left( \begin{array}{ccccc}  
                  0&\quad 2n&\quad 0&\quad &\\
                  0&0&2n-1&&\\
                   &.&.&\ddots&  \\
                 &  & &0&1\\
                 & & & &0   \end{array} \right) \quad \quad
 \hbox{ and } \quad 
 g = D  \left( \begin{array}{ccccccc}
                  0&\quad   &\quad  &  &&& \\
                  0&\quad 0& &&&&\\
                   2&\quad 0&\quad  0 &&&&  \\
                   &\quad 6&\quad 0& \quad 0&&&\\
                   & &\quad 12& \quad 0&  0&&\\
                  &   &&\ddots&\ddots&\ddots& \\
                 & & & &2n(2n-1)&0&0    \end{array} \right)
\label{def:dg}
 \end{equation}
 we obtain
  \begin{equation}
     (d + g)^k = \sum_{k_1+k_2 =k} W_{k_1,k_2}
   \end{equation}
 where  $W_{k_1,k_2}$ is a product of $k$ factors with $k_1$ factors equal
 to $d$ and $k_2$ factors equal to $g$, with $k_1+k_2 = k$
 (there are $2^k$ such terms because  the matrices
 $d$ and $g$ do not commute).
  We remark  that $d$ is an upper-diagonal  band  matrix and
  its non vanishing terms are all   on the band located
 at level +1 above the diagonal. Similarly $g$ is a  lower-triangular
 band matrix  and its non vanishing terms are all  on the band located 2 
levels
 below the diagonal. Therefore a product of $k_1$ matrices $d$ and
   $k_2$  matrices  $g$ will have a non-zero diagonal term only if 
$k_1 = 2 k_2$  i.e. if $k = 3k_2$; hence, $k$  is a multiple of 3.
 For example, we have
  \begin{equation}
 {\rm Tr}(A_{n}^3) = {\rm Tr}(ddg + dgd + gdd) = 3{\rm Tr}(d^2 g)
   = 3D \sum_{l=1}^{2n-1} l(l+1)(2n-l+1)(2n-l) \simeq 
3D (2n)^5 \int_0^1 x^2(1-x)^2 dx =
 D \frac{16n^5}{5} \, .
 \end{equation}
 More generally, the terms that contribute to
  ${\rm Tr}(A_{n}^{3k})$ are obtained by taking the product of
  $2k$ factors $d$ and $k$ factors $g$
 written in all possible orders (there  are
 $\frac{ (3k)!}{(2k)! k!}$  such terms):
  \begin{equation}
   {\rm Tr}(A_{n}^{3k}) = \sum  {\rm Tr}( W_{2k,k} ) =
   d^{2k}g^k + \hbox{`permuted terms'}
    \end{equation}
In particular, we  have
 \begin{eqnarray}
 {\rm Tr}(d^{2k}g^k) &=& D^{k}  \sum_{l} l(l+1)(l+2)..(l+2k-1)
 (2n-l+1)(2n-l)...(2n-l -2k +2) \nonumber
  \\  &\simeq&
 D^{k} (2n)^{4k+1}\int_0^1 x^{2k}(1-x)^{2k} dx  =  D^{k} (2n)^{4k+1}
\frac{(2k)!(2k)!}{(4k+1)!} \, .
\end{eqnarray}
  The trace of any term  $W_{2k,k}$   is  given by the same
 expression at the leading order:   indeed, the elements of the
 matrix  $dg$ are of order $n^3$ whereas those of
 the commutator  $[d, g]$ are of order $n^2$.  
The reason is that both $dg$ and $gd$ are triangular with all nonvanishing 
matrix elements one level below the diagonal, of the form $(dg)_{l,l-1}$ 
and $(gd)_{l,l-1}$. In the center of the matrix, $l\approx n$, a 
generic term is of the order of $n^3+O(n^2)$. 
Therefore both $dg$ and $gd$ are dominated by $n^3$, while $[d,g]$, 
given by the difference of two such terms is dominated by the $O(n^2)$ 
corrections.  We know that  any $W_{2k,k}$   differs from   $ d^{2k}g^k$ 
by a finite number  of commutators.  Therefore, ${\rm Tr}(W_{2k,k}) =  
{\rm Tr}(d^{2k}g^k) +$ subleading terms. We thus have
 \begin{equation}
 {\rm Tr}(A_{n}^{3k}) \simeq  \frac{ D^{k} (2n)^{4k+1}}{(4k+1)}
  \frac{ (3k)! (2k)!}{ (4k)! k!}\, .
\label{TrM3k}
\end{equation}
 Rewriting this  trace in terms of the
 eigenvalues of the matrix $A_{n}$, we obtain
\begin{equation}
    \sum_{i =1}^{2n+1} \lambda_{n,i}^{3k}  \simeq
  \frac{ D^{k} (2n)^{4k+1} }{(4k+1)}
  \frac{ (3k)! (2k)!}{ (4k)! k!} \,.
\label{Sumeigenv}
\end{equation}
 Besides, from   equation~(\ref{TrM3k}),  we deduce that
  \begin{equation}
 \frac{\Big[{\rm Tr}(A_{n}^{3})\Big]^k}{{\rm Tr}(A_{n}^{3k})}
  \sim n^{k-1} \, .
 \label{traceratio}
\end{equation}
 This equation  shows that the trace  is not dominated by the largest
 eigenvalue alone: otherwise this `participation'
  ratio would be of  order one. Rather,
 a finite  fraction $\rho$  of the eigenvalues has a scaling behavior
  similar to that  of $\lambda_{{\rm max}}(n)$ and   we can write:
 ${\rm Tr}(A_{n}^{3k}) \sim \rho  n \lambda_{{\rm max}}(n)^{3k}\,.$
 Thus, we  deduce from
  equation~(\ref{Sumeigenv})  the following behavior
\begin{equation}
     \rho  n  \,  \lambda_{{\rm max}}(n)^{3k} \simeq
  \frac{ D^{k} (2n)^{4k+1} }{(4k+1)}
  \frac{ (3k)! (2k)!}{ (4k)! k!} \,.
\label{Maxeigenv}
\end{equation}
 Finally, using Stirling formula, we obtain for large $k$, assuming $\rho 
n$ does not vary strongly with $n$
  \begin{equation}
  \lambda_{{\rm max}}(n) \simeq D^{1/3}  (2n)^{4/3}
 \Big(   \frac{ (3k)! (2k)!}{ (4k)! k!}    \Big)^{1/3k} \simeq
  \frac{3}{4} D^{1/3}  (2n)^{4/3}  \, .
  \end{equation}
 This ends the proof of equation~(\ref{eq:scaling}) or 
(\ref{eq:scalingIIB}) in the case $\omega =0$.
 In  Table~I  
the  highest eigenvalue of the matrix $A_n$
 is computed numerically for various
  values of $2n$ up to $2n=300$ (for $\omega =0$
 and $D =1$).
 The asymptotic scaling given in equation~(\ref{eq:scaling}) is well
 satisfied.
  
  When  $\omega$ is  different from 0, we can still
 write $A_{n}$ as a sum of two matrices  as in 
equation~(\ref{eq:AnasSum}):
 the  upper-diagonal matrix $d$ remains the same as in 
equation~(\ref{def:dg})  but the  lower-triangular  part $g'$ is  now 
given by the previous  $g$ plus
  a band-diagonal matrix $g_2(\omega)$ (with a band-diagonal at level -1)
  which contains terms proportional  to  $\omega$.
 However, in the large $n$ limit (and keeping the value of $\omega$ fixed)
 the matrix elements of
 $g$  are much bigger than those of  $g_2(\omega)$:
 hence   $g' \simeq g$
 up to subdominant contributions and the large $n$ scaling of the
 maximal eigenvalue is insensitive to   $\omega$   at leading order,
 therefore  equation~(\ref{eq:scaling}), derived for $\omega =0$, remains 
true
  for finite values of $\omega$.  In Tables~II and III
 we   give numerical results $\omega  =1$ and  $\omega  =-1$
(taking  $D=1$ for both cases). The scaling behavior, proportional to
  $(2n)^{4/3}$,
 is well satisfied and 
 it   seems  also  that the prefactor $3/4$ remains correct.

\begin{table}  \centering
  \begin{tabular}{r|r|r}
 $2n$    &  \quad \quad
 $\lambda_{{\rm max}}(n)$  \quad \quad   &  \quad
  $\frac{3D (2n)^{4/3}} {4 \lambda_{{\rm max}}(n)}$  \\
    &     &   \\
  \hline   
   10 &  15    &  1.077  \\
   40 & 101    &  1.015  \\
   80 & 257    &  1.006  \\
  100 & 347    &  1.003  \\
  120 & 443    &  1.0021  \\
  140 & 544    &  1.0022  \\
  160 & 650    &  1.0018  \\
  180 & 761    &  1.0016  \\
  200 & 876    &  1.0013  \\
  300 & 1504.7 &  1.0010
 \end{tabular}
\label{Table1}
\caption{Behavior of the dominant eigenvalue in the case $\omega=0$, 
$D=1$.}
\end{table}

\begin{table}  \centering
  \begin{tabular}{r|r|r}
 $2n$   &  \quad \quad
 $\lambda_{{\rm max}}(n)$  \quad \quad   &  \quad \quad
   $\frac{3D (2n)^{4/3}} {4 \lambda_{{\rm max}}(n)}$ \\
     &     &   \\
   \hline
   10 &  13.2    &  1.223   \\
   40 &  95.9    &  1.072   \\
   80 &  248     &  1.043   \\
  100 &  336     &  1.035   \\
  120 &  431     &  1.031   \\
  140 &  530     &  1.028  \\
  160 &  635     &  1.026  \\
  180 &  745     &  1.023  \\
\end{tabular}
\label{Table2}
\caption{Behavior of the dominant eigenvalue in the case $\omega=1$, 
$D=1$.}
\end{table}

\begin{table}  \centering
  \begin{tabular}{r|r|r}
 $2n$   &  \quad \quad
 $\lambda_{{\rm max}}(n)$  \quad \quad   &  \quad \quad
   $\frac{3D (2n)^{4/3}} {4 \lambda_{{\rm max}}(n)}$ \\
    &     &   \\
   \hline
   10 &  18      &  0.900   \\
   40 &  107.5   &  0.952   \\
   80 &  267     &  0.968   \\
  100 &  358     &  0.972   \\
  120 &  455     &  0.975   \\
  140 &  557.5   &  0.977   \\
  160 &  666     &  0.980   \\
  180 &  777     &  0.981   \\
\end{tabular}
\label{Table3}
\caption{Behavior of the dominant eigenvalue in the case $\omega=-1$, 
$D=1$.}
\end{table}

\section{Boundness of the $\bar{c}_n$}   
\label{sec:Appendix2}

Let us assume an expansion (of the initial moments (\ref{asy4}))
\begin{equation}\label{bound1}
M_{n.l}=\sum_{m=0}^{\infty}b_{n,l}^{(n+m)}\epsilon^{2(n+m)}
\end{equation}
for the $l$-th moment of order $n$ as defined after the matrix 
(\ref{formulaAn}).
The $b$'s are bound (by construction).
Let us expand the $c_{n,l}$ of Eq. (\ref{asy4}) in powers of $\epsilon$ as
\begin{equation}\label{bound2}
c_{n,l}=\sum_{m'}\bar{c}_{n,l}^{(m')}\epsilon^{2m'}\, .
\end{equation}
Since only even moments are considered, the powers of $\epsilon$ are even. 
Now we write Eq. (\ref{asy4}) in the form
\begin{equation}\label{bound3}
M_{n.l}=\sum_{n',l'}c_{n',l'}U_{n',l'}^{(n,l)}\, ,
\end{equation}
where we equate the components of the vectors. now we use the property 
(\ref{asy2}) of the eigenvectors $U_{n',l'}^{(n,l)}$, namely 
$U_{n',l'}^{(n,l)}=0$
for $n>n'$, and solve the equation order by order in $\epsilon$.

Take first $n=1$, and leading order in $\epsilon$ 
\begin{equation}\label{bound4}
b_{1,l}^{(1)}=\sum_{l'}\bar{c}_{1,l'}^{(1)}U_{1,l'}^{(1,l)}\, ~~~l=0,1,2\, 
.
\end{equation}
These are three linear equations for the $\bar{c}_{1,l'}$, since the 
eigenvectors $U_{1,l'}^{(1,l)}$ are given.

Next take $n=2$,
\begin{eqnarray}\label{bound5}
b_{1,l}^{(2)}=\sum_{l'=0}^2\bar{c}_{1,l'}^{(2)}U_{1,l'}^{(1,l)}+
\sum_{l'=0}^4\bar{c}_{2,l'}^{(2)}U_{2,l'}^{(1,l)} \\
b_{2,l}^{(2)}=\sum_{l'=0}^2\bar{c}_{1,l'}^{(2)}U_{1,l'}^{(2,l)}+
\sum_{l'=0}^4\bar{c}_{2,l'}^{(2)}U_{2,l'}^{(2,l)} \, .
\end{eqnarray}
These are eight equations for the 
$\bar{c}_{2,l'}^{(2)},\,\bar{c}_{1,l'}^{(2)}$. 

This process can be 
continued to any order in $\epsilon$. The independence of the eigenvectors 
implies that finite solutions for the $\bar{c}_{n,l}^{(m')}$ can be 
obtained by Kramer's rule. Note, that $b_{n,l}^{(m')}=0$ for $n>m'$,
therefore, also $\bar{c}_{n,l}^{(m')}=0$ for $n>m'$. Therefore,
\begin{equation}\label{bound6}
c_{n,l}\sim\bar{c}_n\epsilon^{2n}
\end{equation}
for small $\epsilon$.

This shows that the $\bar{c}_n$ are bounded for any finite $n$, but it 
does not imply the existence of a uniform bound for all $n$ and $m$.

\end{document}